# PROBLEM AND SOLUTION WITH THE LONGITUDINAL TRACKING OF THE ORBIT CODE

Linhao Zhang[1, 2, 3*], Jingyu Tang[1, 2, 3†], Yukai Chen[1, 3]
[1]Institute of High Energy Physics, Chinese Academy of Sciences, Beijing 100049, China
[2]University of Chinese Academy of Sciences, Beijing 100049, China
[3]Spallation Neutron Source Science Center, Dongguan 523800, China

*Abstract*

The ORBIT code has been widely used for beam dynamics simulations including injection and acceleration in high-intensity hadron synchrotrons. When the ORBIT's 1D longitudinal tracking was employed for the acceleration process in CSNS/RCS, the longitudinal emittance in eV-s was found decreasing substantially during acceleration, though the adiabatic condition is still met during this process. This is against the Liouville theorem that predicts the preservation of the emittance during acceleration. The recent machine study in the accelerator and the simulations with a self-made code demonstrate that the longitudinal emittance is almost invariant, which further indicates that the ORBIT longitudinal tracking might be incorrect. A detailed check-over in the ORBIT code source finds that the longitudinal finite difference equation used in the code is erroneous when applied to an acceleration process. The new code format PyORBIT has the same problem. After the small secondary factor is included in the code, ORBIT can produce results keeping the longitudinal emittance invariant. This paper presents some details about the study.

## INTRODUCTION

The ORBIT code [1] is a particle tracking code written by C++ and based on the SuperCode driver shell. It has been widely used for beam dynamics simulations including injection and acceleration in high-intensity hadron synchrotrons such as the SNS accumulator ring, Fermilab Booster, ISIS/RCS and CSNS/RCS. A new format PyORBIT, based on the Python language, has been developed over the past decade, which is a new implementation and extension of algorithms of the original ORBIT code [2].

The accelerator complex of China Spallation Neutron Source (CSNS) comprises an 80 MeV H- Linac and a rapid cycling synchrotron (RCS). The RCS accelerates the proton beam to 1.6 GeV from 80 MeV within 20 ms. Then the beam is extracted and transferred to the target to produce neutrons. Since Februray 2020, the accelerator has been providing a stable beam of 100 kW, which is the design goal of the CSNS Phase-I project.

In recent machine studies, we found that the measured bunch length during acceleration is inconsistent with the ORBIT simulations. In this paper, the possible causes of this problem are screened and analyzed, and the corresponding solution is presented.

## PHENOMENA ON THE LONGITUDINAL EMITTANCE WITH ORBIT SIMULATION AND MACHINE STUDIES

Following up the early theoretical study on the short bunch extraction in an RCS [3], the machine studies on the bunch compression were conducted recently in the CSNS/RCS. The RMS bunch length during acceleration was found inconsistent with the one obtained from the ORBIT's 1D longitudinal simulation, as shown in Fig. 1. However, the experiment agrees well with the theory that the longitudinal emittance preserves during acceleration, as expressed by:

$$\sigma_{\tau,rms} = \sqrt{\frac{A_{rms}}{\omega_0}} \left( \frac{2|\eta|}{\pi h e V_0 \beta^2 E |\cos\phi_s|} \right)^{1/4}, \qquad (1)$$

where $A_{rms}$, $\omega_0$, $\eta$, $h$ and $\phi_s$ denote the RMS longitudinal emittance in eV-s, revolution frequency of the reference particle, phase-slip factor, harmonic number, and synchronous phase, respectively, $V_0$ is the RF voltage, $\beta$ is the relativity velocity factor, $E$ is the total beam energy.

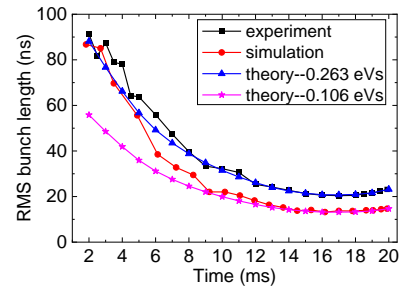

Figure 1: The evolution of the RMS bunch length obtained from the experiment (black squares), the ORBIT simulation (red circles), theoretical calculation with the RMS bunch area of 0.263 eVs (blue triangles) and 0.106 eVs (pink stars) during the CSNS/RCS acceleration from 2 ms to 20 ms.

From Fig. 1, the longitudinal emittance with the ORBIT simulation decreases from 0.263 eVs to 0.106 eVs, which causes puzzling. Figure 2 shows the reduction of the longitudinal emittance in a more direct way. The same problem also exists in the previous studies with the ORBIT code [4, 5] for both 1D and 3D simulations, but has been unresolved for a long time.

___________________
* zhanglinhao@ihep.ac.cn
† tangjy@ihep.ac.cn

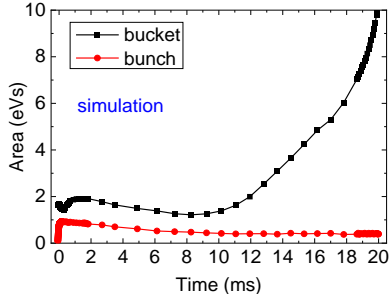

Figure 2: The evolution of bunch area and bucket area during acceleration obtained from the ORBIT simulation.

As predicted by the Liouville theorem, the longitudinal emittance during acceleration remains preserved under the adiabatic condition, which can be estimated by [6]:

$$\alpha_{ad} = \left| \frac{1}{\omega_s^2} \frac{d\omega_s}{dt} \right| \ll 1, \quad (2)$$

where, $\omega_s$ is the angular synchrotron frequency and $\alpha_{ad}$ is the adiabatic coefficient. The adiabatic coefficient during CSNS/RCS acceleration is less than 0.05, as shown in Fig. 3, which means the adiabatic condition is well satisfied, so it should conform to the Liouville theorem. Thus, the longitudinal emittance using the ORBIT 1D simulation is inaccurate here.

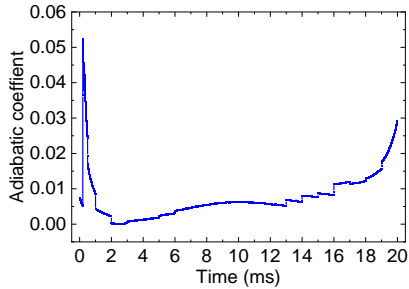

Figure 3: The adiabatic coefficient during CSNS/RCS acceleration, which is almost always less than 0.05.

## ANALYSIS TO THE PROBLEM OF THE LONGITUDINAL EMITTANCE REDUCTION IN THE ORBIT SIMULATIONS

### Simulation conditions

Different simulation conditions in CSNS/RCS are considered and tested to analyze why the longitudinal emittance damps during acceleration with the ORBIT 1D longitudinal tracking.

First, we checked the influence of space charge effects. As shown in Fig. 4(a), the longitudinal emittance damps during acceleration whether the simulation involves the space charge effects or not. Next, an initial longitudinal distribution that neglects the injection painting process and perfectly matches the initial RF bucket is applied to the acceleration. The situation remains the same, see Fig. 4(b). Then, the constant RF bucket area and small bunch area are used, respectively, and it gives the same conclusion.

To find a clue for this problem, we went through detailed checking over the ORBIT source file and found that the problem was caused by the inappropriate use of the longitudinal difference equations, which will be explained in detail below.

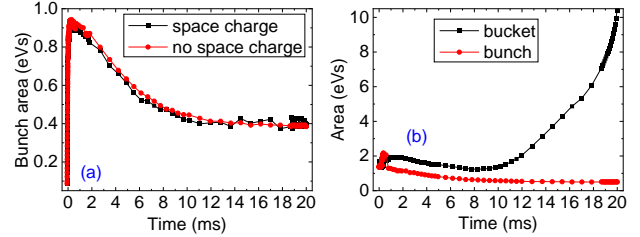

Figure 4: Simulations on the longitudinal emittance during CSNS/RCS acceleration with the ORBIT 1D longitudinal tracking: (a) with/without space charge effects; (b) initially matched bunch distribution.

### Longitudinal equations of motion in ORBIT

The phase of an abitrary particle with respect to the RF voltage can be expressed by [7]:

$$\phi(t) = \int \omega_{RF}(t) \cdot dt - h\theta(t), \quad (3)$$

where $\omega_{RF}(t)$ is the RF angular frequency, $\theta(t)$ is the azimuthal angle of the particle. Thus, when the particle traverses the RF cavity at the $(n+1)$th turn, the corresponding RF phase is:

$$\phi_{n+1} = \phi_n + \omega_{RF,n+1} T_{n+1} - 2\pi h, \quad (4)$$

where $T_{n+1} = C_{n+1} / v_{n+1}$ is the revolution period of the particle at the $(n+1)$th turn with $C_{n+1} = (1 + \alpha_0 \delta_n) C_0$ with $\alpha_0$ the momentum compaction factor, $C_0$ the ring circumference and $\delta_n$ the relative momentum deviation of the particle from the reference particle at the $n$th turn, and $v_{n+1} = c\sqrt{1 - (E_0/E_n)^2}$, with $c$ the speed of light, $E_0$ the rest energy of the particle and $E_n$ the total energy of the particle at the $n$th turn. After the $(n+1)$th transversal of the RF cavity, the total energy of the particle becomes:

$$E_{n+1} = E_n + eV_{0,n+1} \sin \phi_{n+1}. \quad (5)$$

Thus, Eq. (4) and Eq. (5) composes the difference equations of longitudinal motion for any particle in a synchrotron. They can be applied for any particle of a bunch when not considering the intensity effects like the space charge effect.

Equtions (4) and (5) can be rewritten as the more commonly used finite difference equations concerning the RF phase and the energy difference between any particle and the synchronous particle:

$$\delta\phi_{n+1} = \delta\phi_n + \omega_{RF,n+1} T_{n+1} - 2\pi h - (\phi_{s,n+1} - \phi_{s,n}), \quad (6)$$

$$\delta E_{n+1} = \delta E_n + eV_{0,n+1} (\sin \phi_{n+1} - \sin \phi_{s,n+1}). \quad (7)$$

Here, the synchronous phase $\phi_s$ can be obtained below, assuming that the RF voltage has a sinusoidal form:

$$V_0 \sin \phi_s = C_0 \rho \dot{B}, \quad (8)$$

where $\rho$ is the curvature radius of the dipole magnets and $\dot{B}$ is the time derivative of the dipole magnetic field.

Besides, the difference equations of longitudinal motion can also be derived from the equation of a single particle in phase space [7]:

$$\frac{d}{dt}\left(\frac{\delta E}{\omega_0}\right) = \frac{eV_0}{2\pi}(\sin\phi - \sin\phi_s), \quad (9)$$

$$\frac{d}{dt}\delta\phi = h\omega_0\eta\frac{\delta E}{\beta^2 E}. \quad (10)$$

It is worth pointing out that any correct derivation of Eq. (9) must consider the electromagnetic force induced by a varying magnetic field, which is also called the betatron effect [7, 8]. Thus, the corresponding finite difference equations are:

$$\delta\phi_{n+1} = \delta\phi_n + 2\pi h\eta_n\frac{\delta E_n}{\beta_n^2 E_n}, \quad (11)$$

$$\delta E_{n+1} = \frac{\omega_{0,n+1}}{\omega_{0,n}}\left[\delta E_n + eV_{0,n+1}(\sin\phi_{n+1} - \sin\phi_s)\right]. \quad (12)$$

A self-made 1D longitudinal tracking code based on the two sets of finite difference equations, Eq. (6, 7) and Eq. (11, 12), is written and applied to the CSNS/RCS simulation. The results demonstrate that with both sets of equations the longitudinal emittance is invariant within numerical errors, see Fig. 5. Actually, they both consider the betatron effect, which is reflected by changes in the RF frequency or revolution frequency that is synchronized with the varying magnetic field during acceleration. This is particularly important for low to medium energy proton or ion accelerators where there is a large RF frequency ramping. CSNS/RCS is the case.

However, the difference equations for longitudinal tracking in ORBIT adopt the combination of Eq. (7) and Eq. (11). This means that the RF frequency swing during acceleration is not included, thus it results in the spurious longitudinal emittance damping when applied to an RCS, as shown in Figures 1 and 2.

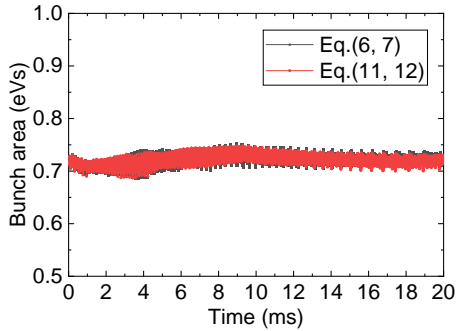

Figure 5: A self-made 1D longitudinal tracking code based on the two sets of finite difference equations, Eq. (6, 7) and Eq. (11, 12), is applied to the CSNS/RCS simulation, which demonstrates that the longitudinal emittance is almost invariant.

## SIMULATION WITH THE REVISED ORBIT

Based on the above analysis, we changed the finite difference equation concerning the energy from Eq. (7) to Eq. (12) in the ORBIT. The new simulation results show that the longitudinal emittance keeps conserved during acceleration, see Fig. 6(a). The new code format PyORBIT has the same problem of the spurious longitudinal emittance damping and has also been modified.

Now the measured RMS bunch length is consistent with the simulations, see Fig. 6(b).

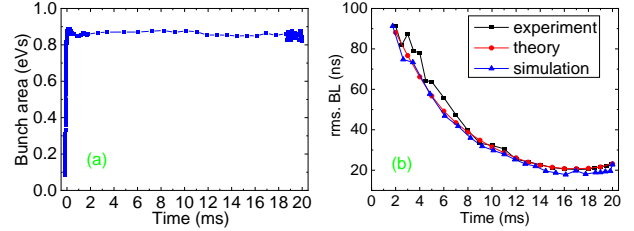

Figure 6: (a) The longitudinal emittance during CSNS/RCS acceleration with the revised ORBIT (including injection); (b) The measured RMS bunch length during CSNS/RCS acceleration (black) is compared with the theoretical calculation (red) and the revised ORBIT simulation (blue).

## SUMMARY

This paper figures out the confusing problem that the longitudinal emittance with the ORBIT 1D longitudinal tracking damps during acceleration in an RCS, which is against the Liouville theorem. The cause is found to be the improper use of the longitudinal finite difference equation concerning the energy in the ORBIT 1D tracking, which neglects the large RF frequency change (the betatron effect) during acceleration. When the small secondary factor is included in the revised ORBIT, the longitudinal emittance keeps invariant and the RMS bunch length obtained from the ORBIT simulation agrees well with the measured one.